\documentclass[aps,prl,twocolumn,groupedaddress]{revtex4}
\usepackage{graphicx,latexsym}

\begin{document}

\title{Transverse Vortex Dynamics in Superconductors}


\author{J. Lefebvre, M. Hilke, R. Gagnon and Z. Altounian}

\affiliation{ Dpt. of Physics, McGill University, Montr\'eal, Canada
H3A 2T8.}

\begin{abstract}

We experimentally characterize the transverse vortex motion and
observed some striking features. We found large structures and peaks
in the Hall resistance, which can be attributed to the long-range
inhomogeneous vortex flow present in some phases of vortex dynamics.
We further demonstrate the existence of a moving vortex phase
between the pinned phase (peak effect) and the field induced normal
state. The measurements were performed on $NiZr_2$ based
superconducting glasses.

\end{abstract}

\maketitle

Type II superconductors placed in a magnetic field (B) will allow
quantized magnetic fluxes to penetrate and form vortex lines
parallel to the field surrounded by superconducting currents.
Because of the sign of these currents, single vortices will repel
each other and condense at zero temperature into an Abrikosov vortex
lattice \cite{Abrikosov57} in the absence of disorder. When
introducing disorder and a driving force, the vortex structure will
evolve through several different phases, which include a moving
Bragg glass, a pinned disordered phase and a liquid-like phase
\cite{GiamarchiPRB57,OlsonPRL81,FangohrPRB64}. Theoretically, it is
expected that the transverse motion of vortices (perpendicular to
the driving force) also exhibits interesting pinning properties,
however these have been elusive to experiments so far.

To overcome the inherent difficulty in observing the vortex flow at
high vortex velocities and densities, where no direct imaging
technique can be used, we used dissipative transport in a very clean
and isotropic type II superconductor described below. In the mixed
state of type II superconductors, the appearance of a resistance is
due to the motion of vortices, which upon application of a current
(J) in the sample will travel in the direction of the Lorentz force
$\vec{J}\times\vec {B}$, thereby inducing a measurable resistance. \
If the vortices move precisely in the direction of the Lorentz
force, that is perpendicular to the current direction, no Hall
voltage is expected. \ Therefore, the condition for the onset of a
Hall voltage is that the vortices be traveling at some angle to the
Lorentz force; then the component of motion parallel to the applied
current will induce a Hall voltage. \ Interestingly, the Hall effect
in the superconducting state still eludes the research community; it
remains controversial even after over 40 years of research on the
subject. \ Some predict a Hall sign reversal below $T_{c}$ caused by
pinning effects \cite{WangPRL67, ZhuPRB60}, others argue that the
anomaly cannot be due to pinning \cite{VinokurPRL71, HagenPRB41,
LobbAS2, BhattacharyaPRL73}, whilst others even predict no sign
reversal at all \cite{BardeenPR140, SchindlerPR130}. \ Moreover the
few studies, which report Hall effect measurements on samples which
also exhibit the peak effect in longitudinal transport measurements
do not show any sharp features \cite{BhattacharyaPRL73, JingPRB42,
NiessenPL15, NiessenJAP40} and no correlation to the different
vortex phases was observed.

Many difficulties involved in the analysis of the Hall resistance
data and theory stem from the competing contributions due to the
Hall resistance of normal electrons and the voltage produced by the
moving vortices. The contribution to the Hall voltage of the
non-superconducting or normal electrons can be found in the vortex
cores as well as in possible pockets of normal phases in an
inhomogeneous superconductor. In order to avoid this problem, we
have chosen a metallic glass, where the Hall voltage contribution of
the normal electrons, antisymmetric in $B$, is negligible compared
to the voltages produced by moving vortices, which is mainly
symmetric in $B$. Indeed, in the normal phase of our system we find
$R_H^{Asy}\simeq B/ne<10 \mu\Omega/T$, where $n>1.4\times10^{22}$
$cm^{-3}$ is the lower bound for the measured electronic density and
$R_H^{Asy}$ is always negligible compared to all other
contributions. These density values are consistent with those found
in for melt-spun NiZr$_{2}$ ribbons \cite{CochranePRB27}.

The measurements of the Hall resistances were performed in glassy
Fe$_{x}$Ni$_{1-x}$Zr$_{2}$ ribbons for different values of $x$ as a
function of magnetic field. The superconducting transition
temperature T$_{c}$ of these high-purity Fe-Ni-Zr based
superconducting metal glasses prepared by melt-spinning
\cite{AltounianJAP53} is around 2.3 K depending on the iron content.
The amorphous nature of the samples ensures that the pinning is
isotropic and has no long-range order, as opposed to crystals, in
which long-range order provides strong collective pinning. Also, due
to their high purity, the samples have a very weak pinning potential
and critical current densities ($J_{c}\leq0.4 A/cm^{2}$) from 10 to
1000 times smaller than in previous typical materials
\cite{HagenPRB41, LobbAS2, BhattacharyaPRL73, JingPRB42,
NiessenJAP40}. \ The advantage of using samples with such a small
depinning current resides in the possibility of investigating the
pinning and depinning mechanisms of the flux line lattice without
the use of a large excitation current which can introduce
uncertainties due to the self-heating it produces. The different
length scales characterizing our superconducting samples were
estimated from standard expressions for superconductors in the dirty
limit \cite{KesPRB28}, and found to be typical of strong type II low
temperature superconductors, as described in ref. \cite{HilkePRL91}.

\begin{figure}
\begin{center}
\vspace*{2cm}
\includegraphics[
width=3.5in]%
{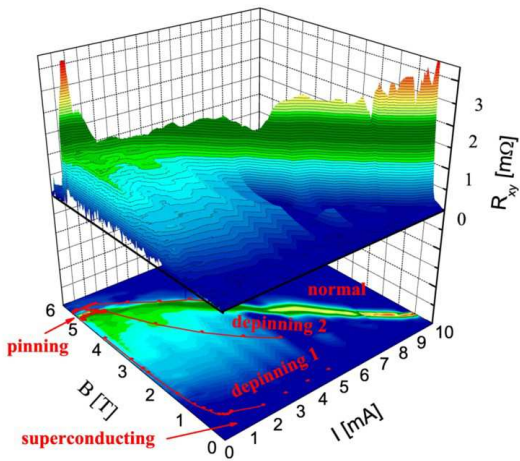}%
\vspace*{-2cm} \caption{The lower projection in red is the phase
diagram obtained from the longitudinal resistance. The color map as
a function of $B$ and $I$ represents the value of the Hall
resistance which is also plotted as a 3d mesh plot. The labeling in
red of the different
phases are extracted from the longitudinal resistances.}%
\label{3D}%
\end{center}
\end{figure}

\begin{figure}
\begin{center}
\includegraphics[
natheight=7.345700in, natwidth=6.355500in, height=4.0888in,
width=3.5405in
]%
{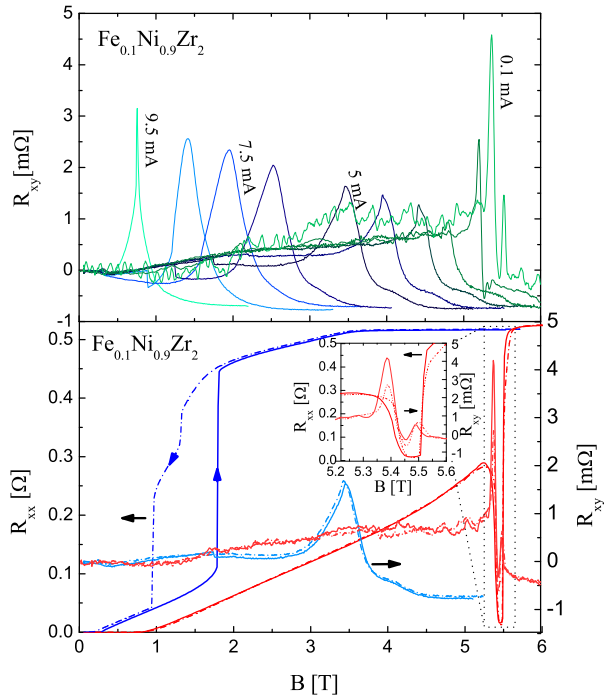}%
\caption{Upper panel: \ Hall resistance as a function of magnetic
field for the following driving currents: 0.1, 1, 2.5, 3.33, 4.16,
5, 6.66, 7.5, 8.3, 9.5, 10 mA. \ Lower panel: \ Longitudinal
resistance and Hall resistance as a function of magnetic field for
up (solid lines) and down (dotted lines) $B$-sweeps with I=0.5mA
(red curves) and I=5mA (blue curves). \ Inset: \ Enlargement of peak
effect region. The experiments were performed in a $^{3}$He system
and most of the data presented here was acquired at temperatures
around 0.4 K. For the resistance measurements an AC resistance
bridge was used at a frequency of 17Hz. The non-zero Hall resistance
above the critical field is due to the small unavoidable
misalignment of the Hall contacts.}

\label{hallE1}%
\end{center}
\end{figure}

In the bottom of figure 1, we present a phase diagram (in red)
obtained from longitudinal resistance measurements for different
driving currents $I$ on a sample of Fe$_{0.1}$Ni$_{0.9}$Zr$_{2}$.
The labeling follows the scheme proposed in ref.\cite{HilkePRL91},\
where the first depinned vortex phase, labeled depinning 1, is
characterized by collective moving vortices and was identified in
ref. \cite{GiamarchiPRB57} as the moving Bragg glass, in which quasi
long-range order exists. At higher $B$, the vortices are pinned
again (pinning phase), which is the origin of the peak effect and
was proposed to originate from the softening of the vortex lattice
\cite{pippard}, which causes the vortex lattice to adapt better to
the pinning potential, or from the destruction of long range order
by disorder described in the collective pinning theory of Larkin and
Ovchinnikov \cite{LO}. \ Finally, just below $B_{c2}$ and for higher
driving currents, an additional depinned vortex phase is observed
(depinning 2) which results from a sudden depinning of the vortex
lattice before the transition to the normal state, and is
characterized by a smectic or plastic flow of vortices
\cite{GiamarchiPRB57, OlsonPRL81, FangohrPRB64}. \ The onset of this
phase is identified in $R_{xx}$ vs $B$ data as the abrupt increase
in resistance following the depinning 1 phase for high driving
currents. For low driving currents the nature of the transition
between the disordered pinned phase and the normal state was never
established, but we show it here to be separated by a depinned phase
as evidenced by the existence of a pronounced peak in the Hall
resistance.

Also shown in figure 1 are the Hall resistances represented as a
topological mesh and color projection as a function of $B$ and for
different driving currents. Graphed in this manner, the Hall data
can be compared directly to the phase diagram and the relation
between these two types of measurements can be established. It is
important to note that a line accounting for the contact
misalignment was subtracted from the Hall curves in this graph.
Strong peaks or features are observed in the Hall resistance for all
driving current, and are found to be located in the depinning 2
phase close to the transition to the normal state in the phase
diagram. In addition, for driving currents below 1 mA, a second peak
is observed right at the onset of the pinning phase. The individual
Hall resistance curves are shown in figure 2, where the peaks
observed in the Hall signal are found to vary in amplitude and shape
with the driving current.

While we have measured more than half a dozen samples of varying
iron concentration all show very similar features and the results
shown in all the figures are representative of all. In all the
samples there is no single clear cut distinction between the
depinning 1 and 2 phases in terms of the Hall resistance, as opposed
to the longitudinal resistance, where a jump in the resistance
allowed us to determine the boundary. However, the features are
always more pronounced in the depinning 2 phase and are highly
reproducible for different $B$-sweeps, which stands in contrast to
the depinning 1 phase where the smaller features change from sweep
to sweep, indicative of a noisy history dependent behavior. This
behavior of the Hall resistance can be understood in terms of the
nature of the different phases. Indeed, in the depinning 1 phase,
which is reminiscent of a moving Bragg glass, one would expect a
small noisy lateral movement along channels, which depends on the
vortex density \cite{GiamarchiPRB57} and would lead to a noisy
$B$-dependent Hall resistance signal. In the depinning 2 phase on
the other hand, the existence of sharp reproducible peaks in the
Hall resistance can be explained by a long range inhomogeneous
vortex flow such as found in smectic channels, where the orientation
can vary very suddenly, depending on the local disorder
configuration and vortex density. Finally, in the pinned phase no
Hall signal is to be expected, which is indeed what we observe.
Generically, a peak in the Hall signal is a measure of a long-ranged
moving vortex structure, since a short-ranged order would be
averaged out over the sample width.

A critical reader could argue that the features seen in the Hall
resistance are simply due to a long range inhomogeneous current flow
as discussed in ref. \cite{Doornbos}. \ Fortunately, it is possible
to show in our case that most of the signal we measure must come
from the intrinsic vortex motion. Indeed, using a DC current allows
us to separate the different contributions. If the current flow path
would solely determine the Hall voltage, this would imply that
$R_H(I,B)\simeq R_H(-I,B)$ and $R_H(I,B)\simeq R_H(I,-B)$, since the
Hall resistance contribution form the normal carriers is negligible.
In our samples, however, the differences are almost as large as the
values themselves, which therefore excludes a large scale
inhomogeneous current flow as the main source for the Hall
resistance. A similar argument can be made for intrinsic vortex
channels, for which $2R_{odd}^{\pm}=R_H(I,\pm B)-R_H(-I,\mp B)$
would have to be zero because the electric field due to the vortex
flow would be opposite for the paired variables $(I,\pm B)$ and
$(-I,\mp B)$ but with the same vortex flow direction. Indeed, the
vortex flow direction is antisymmetric in $I$ and $B$ but the
electric field produced by the vortex motion is symmetric in $B$ and
antisymmetric in $I$. In general, $R_{odd}^{+}$ represents the
vortex flow contributions originating from one edge and
$R_{odd}^{-}$ contributions originating from the other edge. If
$R_{odd}$ is non-zero, this also implies that the vortex motion
cannot be solely described by pure vortex channeling consistent with
our measurements, that $R_{odd}$ is of the same order as $R_H$ (see
figure 3). Moreover, it turns out that $R^{AC}_H\simeq
R_{even}^\pm$, which is also shown in figure 3. This is the reason
that most of the data shown here is in fact $R^{AC}_H$, which is the
even contribution of the Hall resistance and represents an average
over vortices flowing in opposite directions, hence avoiding
intrinsic edge effects. Finally, this demonstrates that the measured
$R^{AC}_H$ is intrinsically due to lateral vortex motion, which
cannot come from pure vortex channeling nor inhomogeneous current
flow.

We can now analyze the peak effect region of the phase diagram
within this framework and show that indeed, there must exist a
moving vortex phase between the pinning phase and the normal state,
since we observe a sharp peak in the Hall resistance when sweeping
the magnetic field through these regions. Even in the lowest
measured currents this peak appears (figure 4), suggesting that a
different vortex phase with long range inhomogeneous vortex flow
such as a smectic phase exists between the peak effect and the
normal state all the way down to vanishingly small driving currents.
A similar peak is seen in all the samples we have measured and to
the best of our knowledge, this is the first reported evidence for
the existence of a smectic-like phase right before the transition to
the normal state in such a low driving regime. It is interesting to
note that the Hall resistance peak becomes smaller with increasing
temperature before vanishing close to $T_c$, which further confirms
that this peak is not due to an inhomogeneous current flow close to
the superconductor to normal transition but rather a consequence of
a long-ranged transverse vortex flow just before the critical field.

\begin{figure}
\includegraphics[width=3.5in]{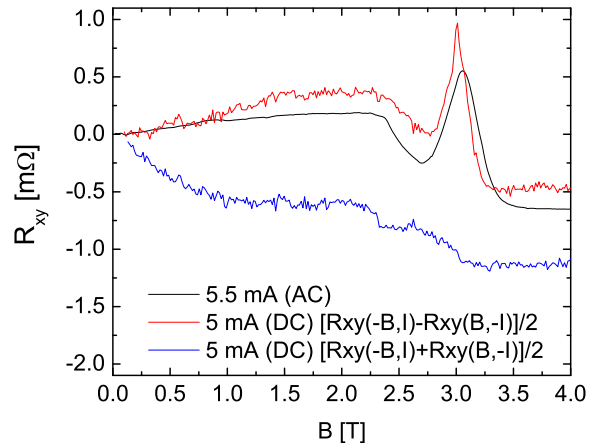}
\caption{Comparison between the Hall resistance obtained with an AC
(17Hz) excitation current compared to
$R_{odd}^{-}=(R_{xy}(-B,I)-R_{xy}(B,-I))/2$ and
$R_{even}^{-}=R_{xy}(-B,I)+R_{xy}(B,-I))/2$ obtained with DC
currents.}
\label{ACDC}%
\end{figure}

\begin{figure}
\begin{center}
\includegraphics[
width=3.3405in
]%
{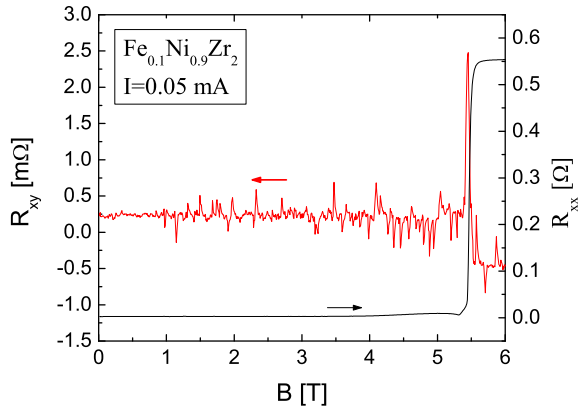}%
\caption{Longitudinal resistance (black curve) and Hall resistance
(red curve) as a function of magnetic field for $B$-sweeps with I=0.05mA.}%
\label{lowestdrive}%
\end{center}
\end{figure}

In summary, we found that in the first depinned vortex phase
encountered as the magnetic field is increased, the Hall resistance
is relatively smooth with small noisy features, which are a result
of some vortices slipping out of the channels in which they flow.
This phase is consistent with a moving Bragg glass. At larger
magnetic fields, the reentrant pinning phase known as the peak
effect, which is characterized by a vanishing longitudinal
resistance also leads to a zero Hall resistance. More interestingly,
at even higher fields and for all driving currents, large features
and peaks are observed in the Hall resistance in the second
depinning phase close to the normal state. These important features
are characteristic of a long range inhomogeneous vortex flow, such
as expected in a smectic phase with orientational changes. Also
important is the strong peak feature observed even at low driving
currents, between the disordered pinned phase and the normal state,
which demonstrates the existence of a long range moving vortex phase
in that region.

\begin{acknowledgments}
The authors acknowledge support from FQRNT and NSERC. Correspondence
and requests for materials should be sent to hilke@physics.mcgill.ca
\end{acknowledgments}

\bibliography{HallPRL}

\end{document}